\documentclass[epj-spec]{svjour}
\usepackage{graphicx}
\usepackage{bm}
\newcommand{\Tdec}{T_\mathrm{dec}}
\newcommand{\Tc}{T_\mathrm{cr}}

\newcommand{\be}[1]{\begin{equation}\label{#1}}
\newcommand{\ee}{\end{equation}}

\usepackage{color}

\begin{document}
\title{Hydrodynamic radial and elliptic flow in heavy-ion collisions
from AGS to LHC energies}
\author{Gregory Kestin \inst{1}\fnmsep\thanks{\email{greg.kestin@gmail.com}.
} 
\and Ulrich Heinz\inst{1,2}\fnmsep\thanks{\email{heinz@mps.ohio-state.edu}.
}}
\institute{Physics Department, Ohio State University, Columbus, 
Ohio 43210, USA 
\and 
CERN, Physics Department, Theoretical Physics Division, CH-1211 Geneva 23, 
Switzerland}
\abstract{
Using ideal relativistic hydrodynamics in 2+1 dimensions, we study the
collision energy dependence of radial and elliptic flow, of the emitted
hadron spectra, and of the transverse momentum dependence of several 
hadronic particle ratios, covering the range from Alternating Gradient 
Synchrotron (AGS) to Large Hadron Collider (LHC) energies. These 
calculations establish an ideal fluid dynamic baseline that can be used
to assess non-equilibrium features manifest in future LHC heavy-ion 
experiments. Contrary to earlier suggestions we find that a saturation 
and even decrease of the differential elliptic flow $v_2(p_T)$ with 
increasing collision energy cannot be unambiguously associated with the 
QCD phase transition.
} 
%
\maketitle
%

\section{Introduction}
\label{sec1}

Relativistic hydrodynamics has proven to be very successful in describing 
the evolution of the hot and dense bulk matter created in heavy-ion 
collisions at the Relativistic Heavy Ion Collider (RHIC) 
\cite{Huovinen:2003fa,Kolb:2003dz,Huovinen:2006jp,experiments}. Since, 
in the ideal fluid limit,
it provides a direct connection between the equation of state (EOS) of
the hot matter and the observed collective flow pattern that can be 
extracted from the measured hadron momentum spectra, this raised the hope
of being able to constrain the QCD equation of state experimentally and
to identify experimental signatures of the quark-hadron phase transition
\cite{Ffm,Rischke95,Kolb:1999it,B00,Huovinen:2005gy} which QCD predicts at 
high temperatures \cite{Karsch:2003jg}. Specifically, ideal fluid dynamics
predicts a non-monotonic collision energy dependence of the $p_T$-integrated
elliptic flow $v_2$ \cite{Kolb:1999it}, with a dip between SPS and RHIC
energies caused by the softening of the QCD EOS around the quark-hadron 
phase transition. Unfortunately, large viscous effects during the late 
hadronic stage of the collision fireball expansion were found to spoil
this phase transition signature \cite{Teaney:2001av,Hirano:2005xf}, and
experimentally the $p_T$-integrated elliptic flow was found to increase
monotonically with collision energy \cite{Alt:2003ab,VoloshinQM06} (see
also \cite{Torrieri:2007qy,Song:2008si}).

Contrary to the $p_T$-integrated elliptic flow, however, the PHENIX 
Collaboration found that the differential elliptic flow $v_2(p_T)$ for
charged hadrons, at two fixed values of $p_T$ ($p_T=0.65$ and 1.75\,GeV/$c$),
does not grow monotonically with increasing collision energy, but
instead appears to saturate at RHIC in the center-of-mass energy range 
between 63 and 200\,GeV/nucleon pair \cite{Adler:2004cj}. In 
Refs.~\cite{Heinz:2004ar,Heinz:2005ja,Lacey:2007na} this observation was 
brought into connection with the QCD quark-hadron phase transition, 
speculating that the non-monotonicity of elliptic flow caused by the 
softening EOS near the phase transition, in spite of being washed out 
in the $p_T$-integrated $v_2$, might survive viscous effects if measured 
at fixed $p_T$. This speculation was based on the observation that the 
monotonically increasing {\em radial} flow, which causes the transverse 
momentum spectra to fall off more slowly at higher collision energies, 
gives bigger weight to the high-$p_T$ region (where $v_2(p_T)$ is larger 
than at low $p_T$) at high than at low collision energies, thereby 
counteracting in the $p_T$-integrated elliptic flow any reduction of 
$v_2(p_T)$ by a softening EOS. Any phase transition signature in the 
integrated $v_2$ should thus manifest itself even more prominently
in the differential elliptic flow $v_2(p_T)$ at fixed $p_T$, and it might
thus remain visible in the energy dependence of $v_2(p_T)$ even in the 
presence of viscous effects that smear out the signature in the 
$p_T$-integrated $v_2$ \cite{Heinz:2004ar,Heinz:2005ja}.

This conjecture remained speculative as long as there existed no systematic 
calculation of the collision energy dependence of the hydrodynamically 
generated differential elliptic flow. The present work fills this hole in
the literature. Our study is based on ideal fluid dynamics even though
the ideal fluid assumption is known to gradually break down below RHIC 
energies \cite{Heinz:2004ar}. We do not attempt to quantitatively describe
relativistic heavy ion data, but to provide theoretical insights into the
systematics of the energy dependence of radial and elliptic flow within
the ideal fluid picture. However, since the validity of the ideal fluid 
picture is expected to improve with increasing collision energy, providing 
an almost quantitative description of most low-$p_T$ phenomena already 
at upper RHIC energies, the calculations presented here can be taken as 
a prediction for hadron spectra and their elliptic flow at LHC energies 
where effects from late hadronic viscosity are expected to become 
negligible \cite{Hirano:2007xd}. As long as the viscosity of the 
quark-gluon plasma (QGP) stage (which at the LHC dominates over the 
hadronic one) remains sufficiently small, these predictions should give 
an accurate description of LHC data\footnote{As the beam energy and 
  thus the initial fireball temperature $T$ increases,
  the effective coupling strength $\alpha_\mathrm{s}(T)$ decreases 
  logarithmically; hence, the shear viscosity to entropy ratio $\eta/s$ is
  expected to slightly increase from RHIC to LHC, suggesting stronger
  dissipative effects at the LHC. However, the particle density $n\sim T^3$ 
  and mean free path $\lambda \sim 1/T$ vary much more rapidly with $T$; 
  correspondingly, the sound attenuation length $\Gamma_\mathrm{s}=\eta/(sT)$ 
  decreases from RHIC to LHC, and at any given time $\tau$ viscous effects, 
  characterized by the ratio 
  $\frac{\Gamma_\mathrm{s}}{\tau}=\frac{\eta}{s}\frac{1}{T\tau}$, 
  should be weaker at LHC energies than at RHIC.}.
 
In addition to providing ideal fluid dynamical benchmarks for the LHC,
an important finding of the present work is that a saturation and even 
decrease with growing beam energy of the differential elliptic flow 
$v_2(p_T)$ at fixed $p_T$ cannot be unambiguously associated with the
quark-hadron phase transition. We find that, at sufficiently high 
collision energies, $v_2(p_T)$ at fixed $p_T$ {\em decreases} with 
increasing beam energy even when the matter is initially so dense that 
all elliptic flow is generated far above the phase transition and 
therefore not affected by the softening of the EOS near the critical 
temperature $\Tc$. This {\em decrease} of the $p_T$-differential elliptic 
flow at fixed $p_T$ is accompanied by a simultaneous {\em increase} 
of the $p_T$-integrated elliptic flow. It is caused by an {\em increase
of the radial flow} with growing beam energy which pushes the hadrons to
larger $p_T$ and renders the momentum spectra less anisotropic at low
$p_T$. Our finding contradicts earlier speculations \cite{Heinz:2004ar,%
Heinz:2005ja,Lacey:2007na} and makes the task of identifying the QCD 
phase transition experimentally even more difficult than previously 
thought. 

The present study is similar in spirit to and complements recent work
by Niemi and Eskola {\it et al.} who also made ideal fluid dynamical
predictions for hadron spectra \cite{Eskola:2005ue} and elliptic flow
\cite{Niemi:2008ta} at LHC energies. Our analysis goes beyond theirs in
its systematic investigation of the beam energy dependence of these 
hadronic observables.
   
\section{Procedure}
\label{sec2}

Our hydrodynamic simulations were performed with the (2+1)-dimensional 
hydrodynamic code {\tt AZHYDRO} that solves the equations of motion in the
two directions transverse to the beam direction for an ideal fluid 
undergoing boost-invariant longitudinal expansion \cite{Kolb:1999it,%
Kolb:2002ve,AZHYDRO}. To run the simulations we must initialize the 
program by providing the following parameters: 
\begin{enumerate}
\item impact parameter $b$;
\item initial proper time $\tau_{0}$ at which the fluid is considered to 
      be in local thermal equilibrium and the hydrodynamic expansion stage 
      begins;
\item initial peak entropy density $s_{0}$ in central ($b=0$) collisions: 
      this parameter is used to control the final charged hadron multiplicity
      $dN_\mathrm{ch}/dy$ and serves as a proxy for the collision energy 
      (see discussion below);
\item peak value $n_{B,0}$ for the initial net baryon number density 
      in central collisions: this parameter is used to control the
      finally observed anti-proton/proton ratio;
\item equation of state (EOS): we use an EOS that matches an ideal gas of
      quarks and gluons at a critical temperature $\Tc=164$\,MeV to a
      realistic hadron resonance gas with non-equilibrium hadron abundances
      whose chemical composition is frozen in chemical equilibrium at 
      $\Tc$ \cite{Kolb:2002ve,Hirano:2002ds,Teaney:2002aj};
\item freeze-out temperature $\Tdec$: this parameter controls where the
      hydrodynamic stage ends and how much collective flow builds up
      during the expansion. We use $\Tdec=100$\,MeV.
\end{enumerate}
We simulate Au+Au collisions. The initial transverse entropy and baryon 
number density profiles are taken to be proportional to each other 
(constant entropy per baryon) and are calculated from the Glauber model
with a mixture of soft and hard collisions, assuming 75\% weight for
the soft contribution, proportional to the transverse density of wounded
nucleons, and 25\% weight for the hard contribution, proportional to
the transverse density of binary nucleon-nucleon collisions (see 
\cite{Kolb:2003dz} for details). This mixture produces the correct 
centrality dependence of the charged hadron multiplicity at RHIC 
\cite{Kolb:2003dz}, and we assume that it doesn't change appreciably
between RHIC and LHC. For the systematics of our study this is not a 
critical assumption. The initial transverse flow velocity at $\tau_0$ is 
assumed to vanish.

In the absence of shocks, ideal fluid dynamics conserves entropy. Therefore, 
the observed charged hadron multiplicity $dN_\mathrm{ch}/dy$, which measures 
the entropy of the final state, can be directly related to integral over 
the initial entropy density profile, parametrized by $s_0$. Hydrodynamics
cannot predict its own initial conditions, and a fundamental theory that
reliably predicts the charged multiplicity as a function of collision energy
$\sqrt{s}$ does not exist yet. We therefore use $dN_\mathrm{ch}/dy$ or,
equivalently, the initial peak entropy density in central Au+Au collisions, 
$s_0$, as a proxy for the collision energy: At any given collision energy, 
a measurement of $dN_\mathrm{ch}/dy$ in the most central collision events 
fixes the value of $s_0$ to be used in ideal fluid simulations at that 
energy. 

Using linear longitudinal expansion without transverse flow at very early 
times, $dN_\mathrm{ch}/dy$ and $s_0$ are thus related by
\begin{equation}
\label{eq1}
  \frac{dN_\mathrm{ch}}{dy} \propto 
  \tau_0 \int d^2x_\perp\, s(\bm{x}_\perp,\tau_0) \propto \tau_0 s_0.
\end{equation}
This relation involves two constants of proportionality, of which the first 
is the inverse of the average entropy per charged particle in the final state 
and is thus determined by the EOS at freeze-out, while the second depends 
on the shape of the initial entropy density profile which is fixed by the 
employed Glauber model. Equation (\ref{eq1}) shows that $s_0$ and $\tau_0$ 
are inversely proportional to each other. Since we vary the collision energy
over orders of magnitude, the initial energy and entropy densities likewise
vary by large factors. It is therefore not adequate to assume a constant 
thermalization time $\tau_0$ independent of collision energy. Due to larger
densities and temperatures at higher collision energies, thermalization
should happen more rapidly. We assume that $\tau_0$ scales with collision
energy by satisfying a quantum mechanical uncertainty relation between
the initial temperature $T_0$ (i.e. the initial average thermal energy)
and the initial time $\tau_0$: $T_0 \tau_0=\mathrm{const.}$ 
\cite{Kapusta:1992uy}. The constant is fixed by assuming the standard
value $\tau_0=0.6$\,fm/$c$ at RHIC energies, $\sqrt{s}=200$\,GeV/nucleon 
pair (it turns out to be approximately 1 \cite{Kolb:2003dz}). Since for 
all collision energies considered here the fireball center in $b=0$ 
collisions is in the QGP phase, $s_0 \propto T_0^3$. Combining these 
relations with Eq.~(\ref{eq1}), one finds
\begin{eqnarray}
\label{eq2}
  \frac{\tau_0\left(\sqrt{s_1}\right)}{\tau_0\left(\sqrt{s_2}\right)} =
  \frac{T_0\left(\sqrt{s_2}\right)}{T_0\left(\sqrt{s_1}\right)} =
  \sqrt{
        \frac{dN_\mathrm{ch}/dy\left(\sqrt{s_2}\right)}
             {dN_\mathrm{ch}/dy\left(\sqrt{s_1}\right)}
       },
\qquad
  \frac{s_0\left(\sqrt{s_1}\right)}{s_0\left(\sqrt{s_2}\right)} =
  \left(
        \frac{dN_\mathrm{ch}/dy\left(\sqrt{s_1}\right)}
             {dN_\mathrm{ch}/dy\left(\sqrt{s_2}\right)}
  \right)^{3/2}.
\end{eqnarray}
  
%
\begin{figure}[htb]
\begin{center}
\includegraphics[bb=45 60 554 750,height=0.75\linewidth,angle=-90,clip=]%
                {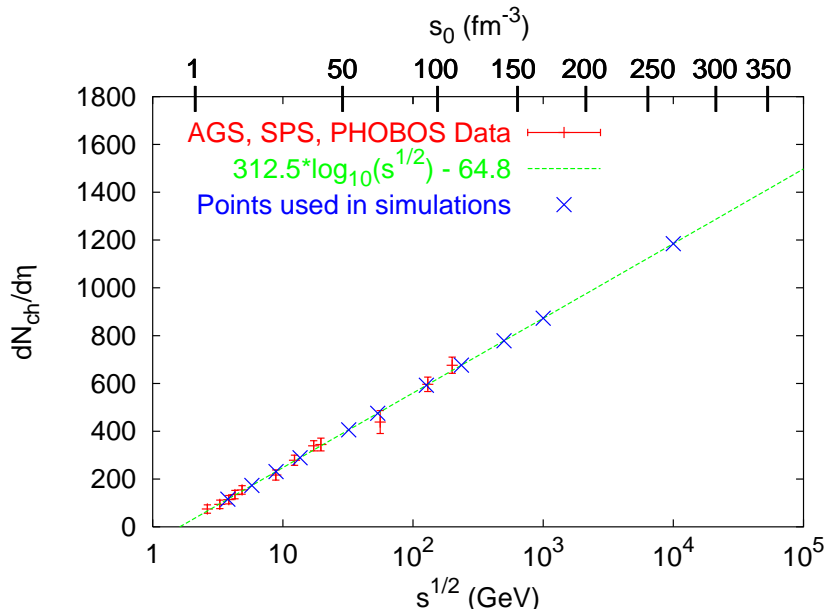}
\end{center}
\caption{(Color online) Charged particle multiplicity $dN_\mathrm{ch}/d\eta$ 
vs. center of mass energy per nucleon pair $\sqrt{s}$ (lower horizontal axis) 
and initial peak entropy density $s_0$ (upper horizontal axis). The 
experimental data are from central Au+Au and Pb+Pb collisions at AGS, SPS, 
and RHIC energies and were compiled in Ref.~\protect{\cite{PHOBOS}}. The 
dashed line is a linear fit to the data. Crosses indicate values for $s_0$ 
and $dN_\mathrm{ch}/d\eta$ for which hydrodynamic simulations were 
performed. See text for details.
}
\label{F1}
\end{figure}
%

Figure~\ref{F1} shows a compilation \cite{PHOBOS} of measured hadron 
multiplicities from ($A{\approx}200$)+($A{\approx}200$) collisions 
at a variety of collision energies explored at the AGS, SPS and RHIC, 
together with a linear fit in $\log_{10}\sqrt{s}$:
\begin{equation}
\label{eq3}
  \frac{dN_\mathrm{ch}}{d\eta} = 312.5 \log_{10}\sqrt{s}-64.8
\end{equation}
The upper horizontal axis maps the multiplicities on the vertical axis
onto $s_0$ values, using Eq.~(\ref{eq2}) with $s_0(\sqrt{s}{=}200\,A\,
\mathrm{GeV})=117$/fm$^3$ \cite{AZHYDRO,Kolb:2003dz}. The linear fit 
(\ref{eq3}) provides guidance for which values to expect for 
$dN_\mathrm{ch}/d\eta$ and $s_0$ at hitherto unexplored LHC 
energies. We emphasize, however, that $s_0$ is not directly related
to $\sqrt{s}$, but only indirectly via Eq.~(\ref{eq1}) through the
charged multiplicity measured at that value of $\sqrt{s}$. For this 
reason we have performed several ``LHC simulations'' with a variety of
$s_0$ values within a reasonable range of the value predicted by the
simple-minded linear extrapolation of existing data shown in Fig.~\ref{F1}.
The first day of LHC heavy-ion experiments will tell us which of these
simulations to choose as hydrodynamic reference for the data.  

We finally comment on our choice of the initial peak net baryon density
$n_{B,0}$. At RHIC energies it was fitted to the measured $\bar p/p$ ratio
to give $n_{B,0}=0.44$/fm$^3$ at $\tau_0=0.6$\,fm/$c$. We did not attempt to
fit the $\sqrt{s}$-dependence of $n_{B,0}$ from $\bar p/p$ ratios measured 
at lower collision energies. Instead we simply held $n_{B,0}=0.44$/fm$^3$ 
constant over the entire energy range while reducing $\tau_0$ with 
increasing collision energy according to Eq.~(\ref{eq2}). 
This implies that, at fixed $\tau$, $n_B$ decreases
with increasing multiplicity as $1/\sqrt{dN_\mathrm{ch}/dy}$. Together 
with the increasing $s_0$ values, this leads to a significant (although 
perhaps not quite strong enough) decrease of the net baryon to entropy 
ratio at midrapidity. Correspondingly, the baryon chemical potential at 
chemical decoupling decreases, reflecting increasing nuclear transparency 
and decreasing baryon stopping power at higher energies. At our highest 
$s_0$ value, $s_0=271$/fm$^3$ (last cross in Fig.~\ref{F1}), we simply 
set $n_{B,0}{\,=\,}0$, assuming approximate baryon-antibaryon symmetry 
at midrapidity at the LHC.

\section{Radial Flow}
\label{sec3}
\subsection{Transverse momentum spectra from AGS to LHC}
\label{sec3a}

\begin{figure*}
\includegraphics[bb=13 28 767 528,width=\linewidth,clip=]{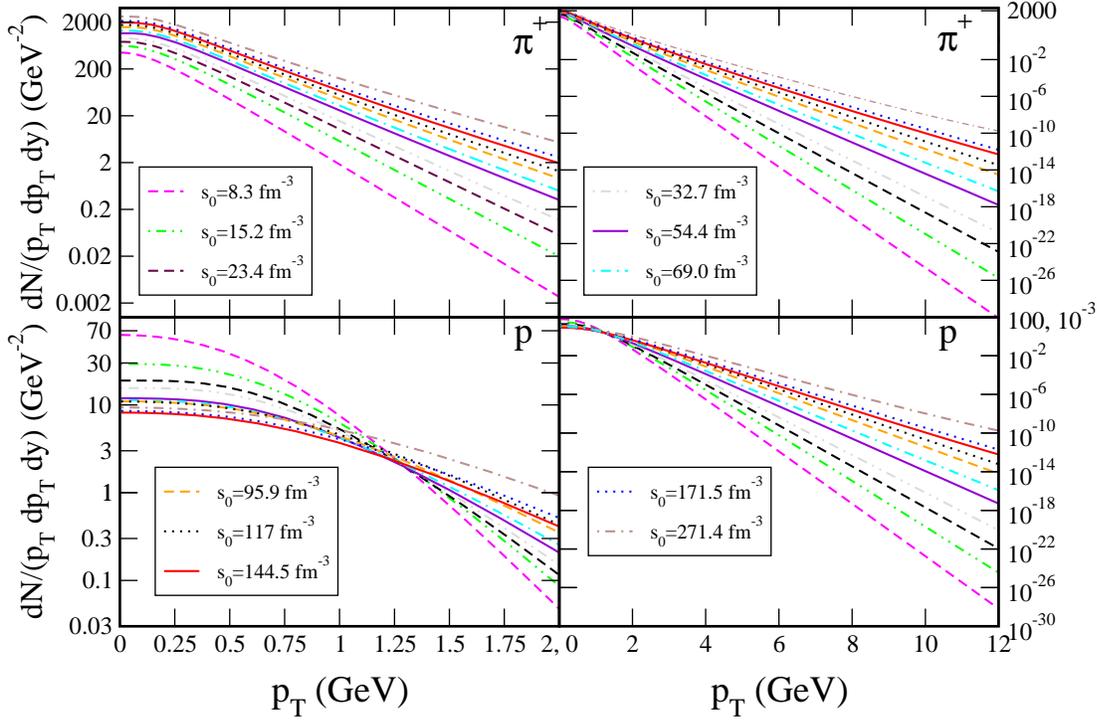}
\caption{(Color online) Transverse momentum spectra for thermal pions
($\pi^+$, upper panels) and protons ($p$, lower panels) at low (left 
panels) and intermediate (right panels) values of $p_T$, for central 
Au+Au collisions with a variety of initial peak entropy densities $s_0$.
\label{F2}}
\end{figure*}

Figure \ref{F2} shows the $p_{T}$-spectra for directly emitted $\pi^{+}$ 
mesons (upper row) and protons (lower row). In this plot we have 
neglected feed-down from resonance decays; its inclusion is 
computationally intensive but, since we keep $\Tdec$ the same at all 
collision energies, it will not qualitatively affect the systematics 
shown in Fig.~\ref{F2}. Here and in the following the curves are labelled 
by the value of $s_0$. The reader can use Fig.~\ref{F1} to translate this 
value into charged hadron multiplicities (which include resonance 
feeddown) and to estimate the corresponding collision energy.

The normalization of the pion $p_T$-spectra is seen to increase with $s_0$, 
reflecting the growth of the total multiplicity with increasing collision
energy. They also become systematically flatter as $s_0$ increases; since
the decoupling temperature $\Tdec$ is held fixed, this is an unambiguous 
signature for larger radial flow at higher collision energies.

%
%

The radial flow effects on pion and proton spectra are similar at large 
transverse momenta (right pnaels in Fig.~\ref{F2}) where their rest mass 
difference can be neglected and the flow effect can be understood in 
terms of the simple blueshift formula $T_\mathrm{slope} = \Tdec
\sqrt{\frac{1{+}\langle v_\perp\rangle}{1{-}\langle v_\perp\rangle}}$
\cite{Schnedermann:1993ws,Concepts}. At low $p_T$, however, they are much
more pronounced for the heavier protons (lower left panel in Fig.~\ref{F2})
\cite{Schnedermann:1993ws,Concepts}. One sees that the radial flow pushes 
the protons away from $p_T{\,=\,}0$, leading to both a dramatic flattening of 
the $p_T$-spectrum and a decrease of the proton yield at low $p_T$, in 
spite of the overall increase in multiplicity. (It should be mentioned 
that the decrease in proton yield at low $p_T$ is amplified by the decrease 
of the baryon chemical potential at higher collision energies, but we 
checked that it is also visible in the antiproton spectra.) Only at the
highest collision energies, where the low-$p_T$ proton spectra are almost
flat, the low-$p_T$ proton yields are seen to follow the general increase
in multiplicity.

\subsection{$\bm{p_T}$- and $\bm{m_T}$-dependence of particle ratios: 
            RHIC vs. LHC}
\label{sec3b}

\begin{figure}[htb]
\includegraphics[width=0.495\linewidth,clip=]{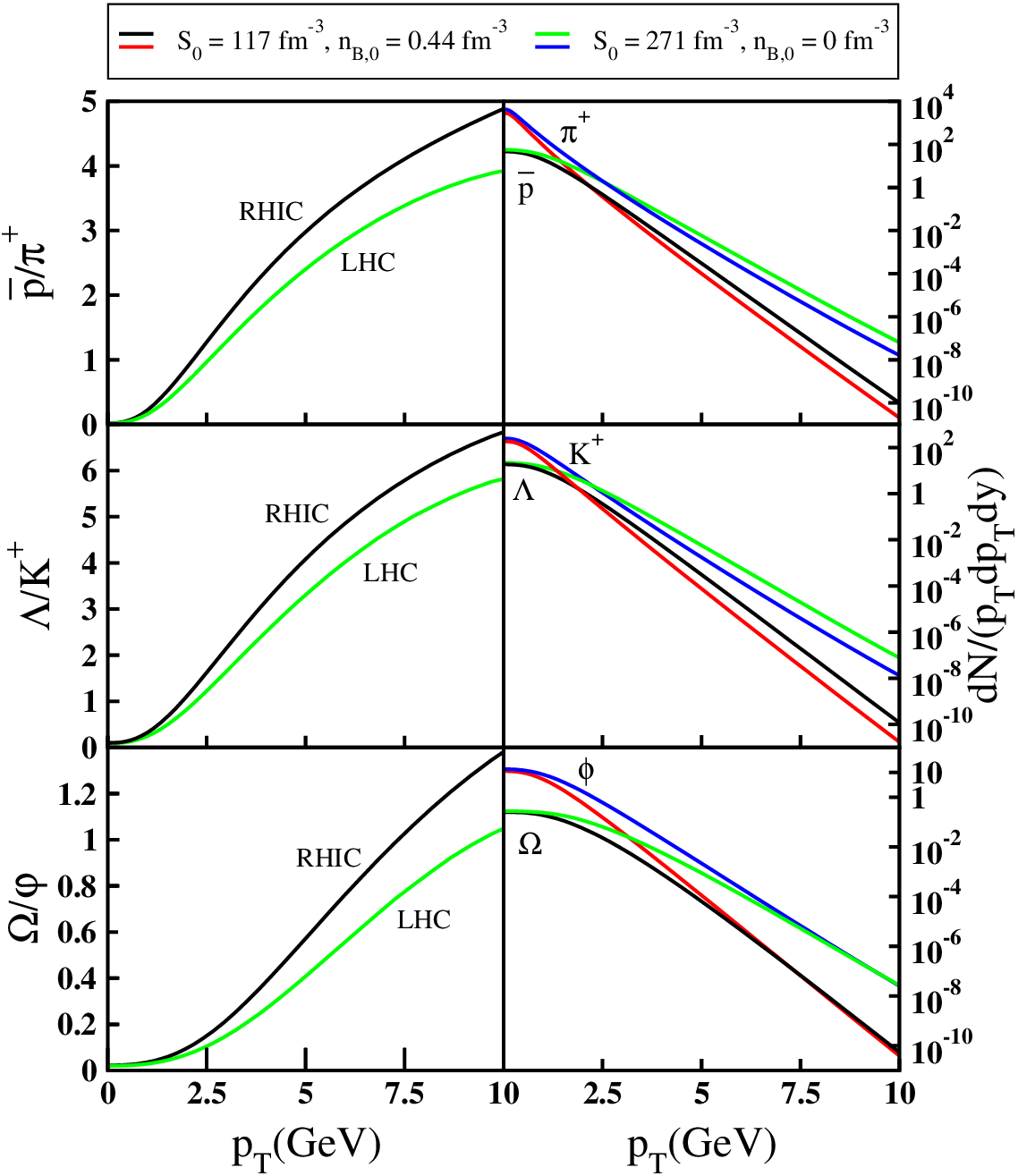}
\includegraphics[width=0.495\linewidth,clip=]{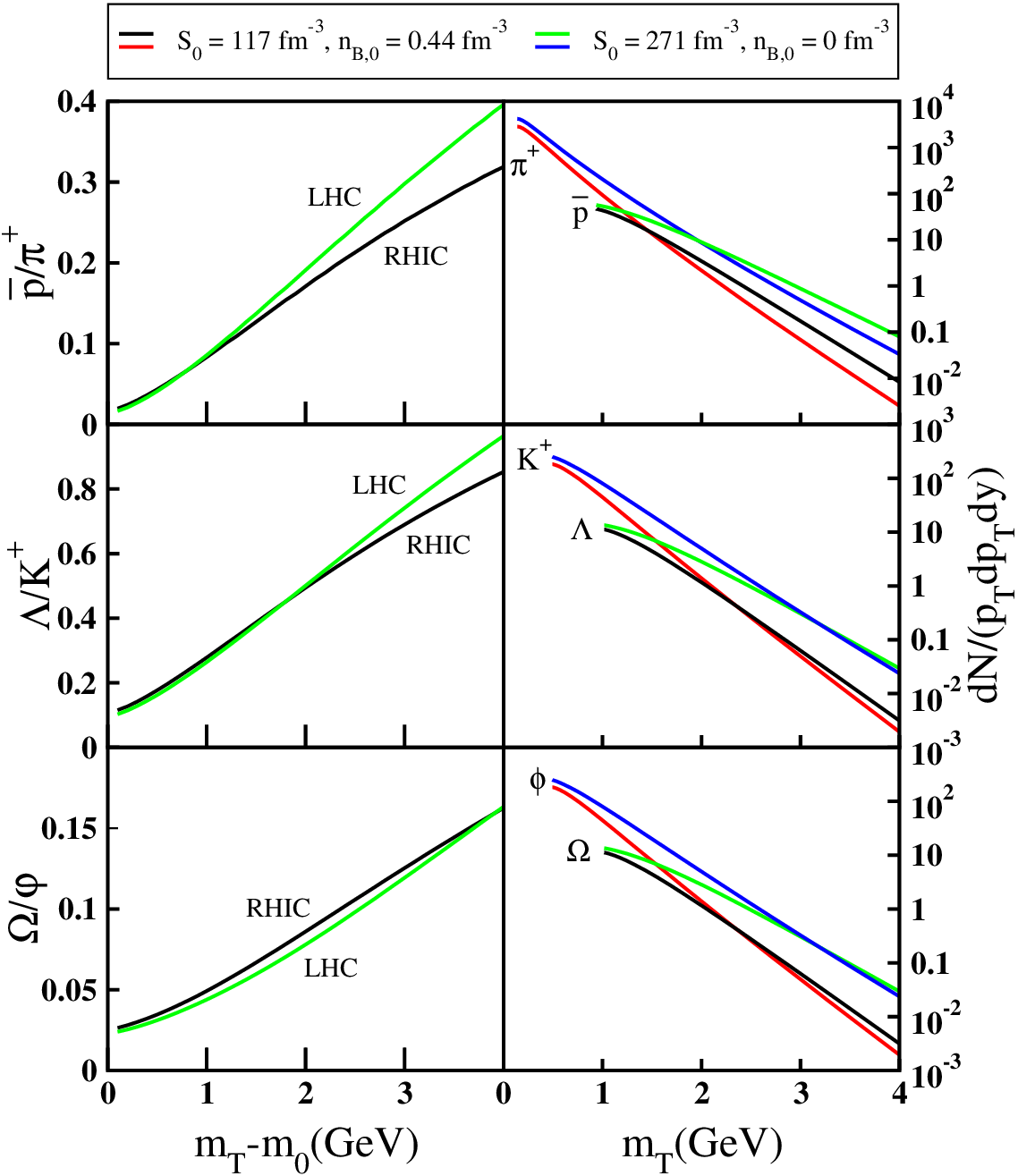}
\caption{(Color online) {\sl Left panel:} Transverse momentum spectra 
(right half) and selected particle ratios (left half) as functions of 
$p_T$ for central Au+Au collisions at RHIC and LHC energies (see text 
for details). A zoomed version of this plot which focusses on the region 
$p_T<2$\,GeV/$c$ can be found in Fig. 53 of Ref. \cite{Abreu:2007kv}.
{\sl Right panel:} Same as left panel, but plotted as a function of 
transverse mass $m_T$ or transverse kinetic energy $m_T-m_0$, 
respectively, over the $m_T$ range where hydrodynamics is expected to be 
a valid description. All curves shown in this Figure include the 
contributions from resonance decays.
}
\label{F3} 
\end{figure}

Hydrodynamic radial flow, which leads to flatter $p_T$-spectra for heavy 
particles, is a key contributor to the observed \cite{experiments} strong 
rise of the $\bar p/\pi$ and $\Lambda/K$ ratios at low $p_T$ at RHIC 
\cite{Kolb:2003dz}. The left column in the left panel of Figure~\ref{F3} 
shows that this rise is predicted to be slower at the LHC than at RHIC.
Since {\em all\,} spectra are flatter at the LHC due to increased radial 
flow (right column of the left panel) while their asymptotic ratios 
at $p_T{\,\to\,}\infty$ (given by their fugacity and spin degeneracy 
ratios \cite{Kolb:2003dz}) remain similar, the ratios between the spectra
increase more slowly with $p_T$.

%
%

In the right panel of Fig.~\ref{F3} we redraw the curves shown in the left
panel as a function of transverse mass $m_T$ (for the spectra) or transverse 
kinetic energy $m_T{-}m_0$ (for the particle ratios). We do so in order to
isolate flow effects, by eliminating the kinematic contribution to the rise 
of the $p_T$-dependent heavy/light hadron ratios that results from plotting 
thermal distributions (which depend on $\frac{m_T}{\Tdec}$) as a function 
of $p_T=\sqrt{m_T^2{-}m_0^2}$. 
In the absence of radial flow, the $m_T$-spectra would show perfect 
$m_T$-scaling, and (except for minor effects from resonance feeddown and 
Bose statistics for pions) the ratios between spectra of particles with 
different rest masses would thus be independent of $m_T$ or $m_T-m_0$. The 
rise of the ratios shown in the left half of the right panel of Fig.~\ref{F3} 
(which is much weaker than that in the left half of the left panel) can 
thus be attributed almost exclusively to radial flow effects. A comparison 
of future LHC data on $(m_T{-}m_0)$-dependent heavy/light particle
ratios with the hydrodynamical predictions shown in Fig.~\ref{F3} can 
therefore help to separate collective flow effects from more exotic 
explanations such as ``baryon junctions'' \cite{Abreu:2007kv}.

\section{Elliptic Flow}
\label{sec4}

While ideal fluid dynamics begins to break down below RHIC energies, due 
to viscous effects in the late hadronic stage which persist even at RHIC 
\cite{Teaney:2001av,Hirano:2005xf}, its validity is expected to improve at 
the LHC where the elliptic flow saturates in the QGP stage and effects from 
late hadronic viscosity become negligible \cite{Hirano:2007xd}. Early 
viscous effects in the QGP stage seem small at RHIC 
\cite{Kolb:2003dz,Hirano:2005xf}, and recent results from Lattice QCD 
indicate little change of its specific shear viscosity $\eta/s$ from RHIC 
to LHC \cite{Meyer:2007ic}. The following {\em ideal fluid} dynamical 
predictions for soft ($p_T < 2{-}3$\,GeV/$c$) hadron production in 
$(A{\approx}200){+}(A{\approx}200)$ collisions at the LHC should thus be 
quite robust.

%
\begin{figure*}
\includegraphics[bb=24 28 746 528,width=\linewidth,clip=]{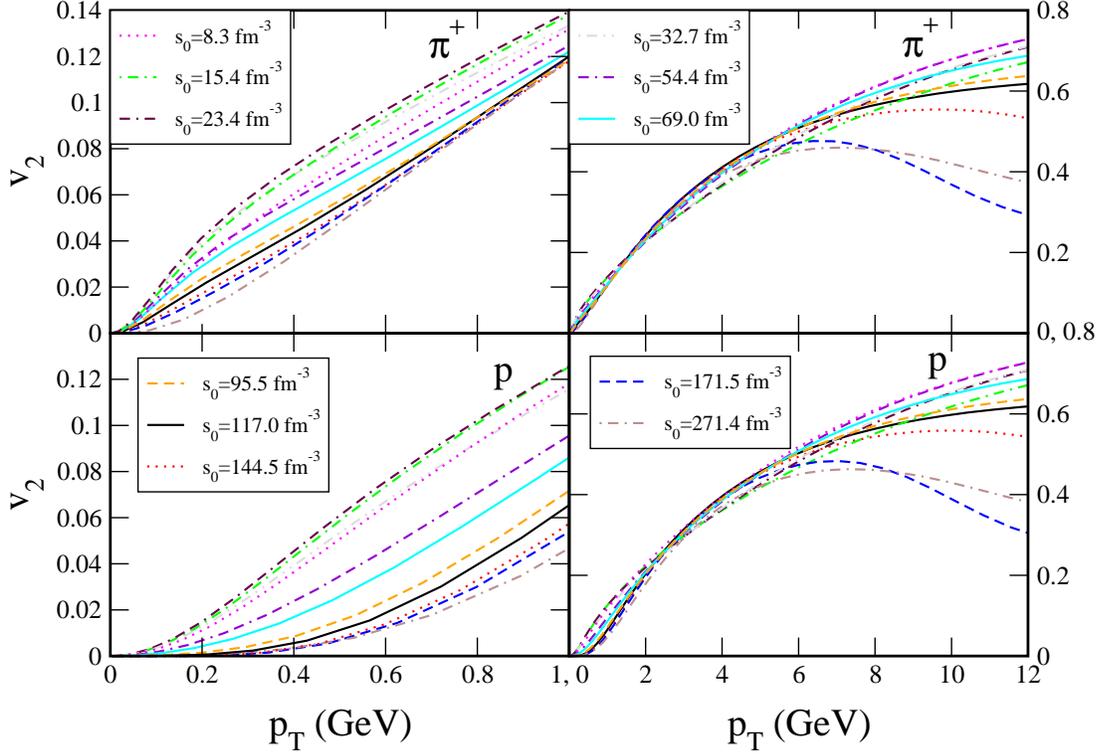}
\caption{(Color online) Differential elliptic flow $v_2(p_T)$ for directly 
emitted pions ($\pi^+$, upper panels) and protons ($p$, lower panels) at 
low (left panels) and intermediate (right panels) values of $p_T$, for 
$b=7$\,fm Au+Au collisions with a variety of initial peak entropy densities 
$s_0$.
\label{F4}}
\end{figure*}
%

In Figure \ref{F4} we look at the differential elliptic flow for both 
protons and $\pi^{+}$ as a function of transverse momentum for $b=7$\,fm 
Au+Au collisions at various collision energies. The left panels are 
expanded views of the low-$p_{T}$ end. At all collision energies, over 
99 percent of all particles are emitted with transverse momenta below 
1.5 GeV/$c$; for this reason we focus our attention on the system's 
characteristics at low $p_T$.\footnote{The astute reader may be puzzled, 
  as we were initially, by the fact that, for the highest collision energies
  studied here, the elliptic flow peaks at intermediate $p_T$ and then
  decreases again, instead of monotonically increasing with $p_T$. This 
  effect appears to be caused by the following phenomenon: even though
  {\em on average} the hydrodynamic flow at freeze-out is stronger in
  the reaction plane than perpendicular to it, at high collision energies
  the {\em largest} value of the flow velocity is found on the freeze-out 
  surface in out-of-plane direction. Since the highest-$p_T$ hadrons are
  emitted from fluid cells with the largest flow velocities, this causes
  a reduction of $v_2$ at very high $p_T$, causing it eventually to even
  turn negative \cite{pasi_private}.
  We thank P. Huovinen for a clarifying discussion on this point.}  

By scanning the curves in order of increasing $s_0$ one notices for both
pions and protons that, at low $p_T$, the differential elliptic flow 
at fixed $p_T$ is not monotonic with collision energy. At the low-energy 
end $v_2(p_T)$ first increases with $s_0$, but when the collision energy 
is further increased beyond $\sqrt{s}\geq10$\,GeV/nucleon pair, the 
differential elliptic flow decreases again.

%
\begin{figure}
\begin{center}
\includegraphics[bb=15 25 543 708,width=0.7\linewidth,clip=]{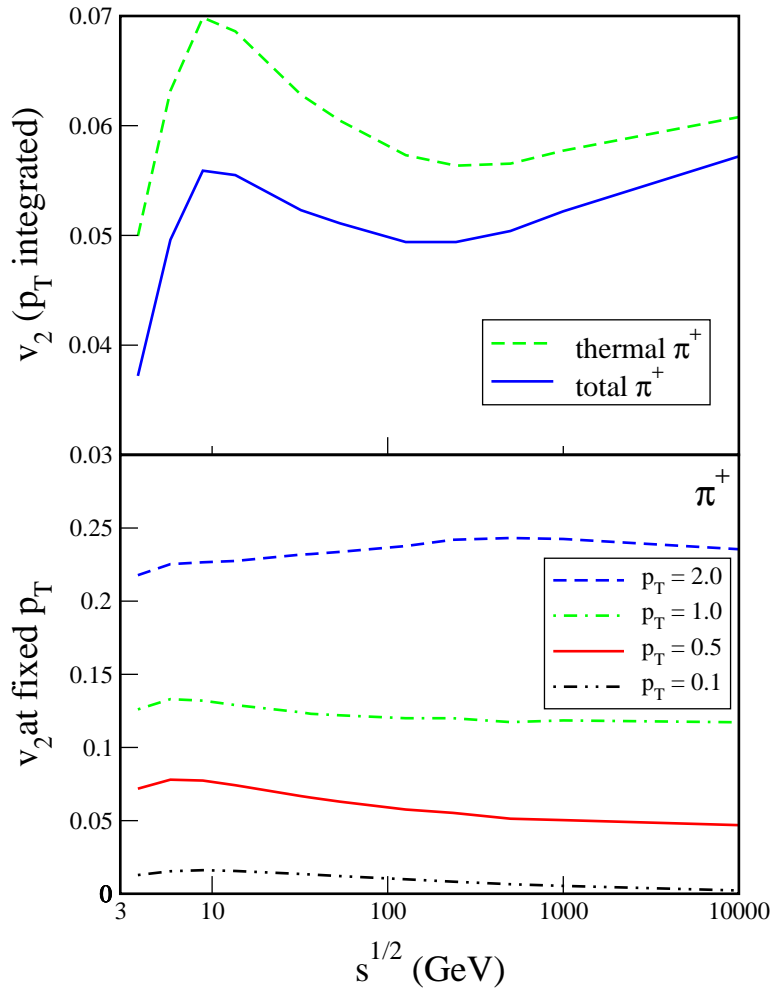}
\end{center}
\caption{(Color online) {\sl Upper panel:} $p_T$-integrated elliptic flow
for pions from $b=7$\,fm Au+Au collisions as a function of collision energy, 
assuming the relation shown in Fig.~\ref{F1} between $s_0$ and $\sqrt{s}$.
The solid line takes only directly emitted thermal pions into account, the
dashed line includes pions from resonance decays. {\sl Lower panel:} Energy 
dependence of thermal pion elliptic flow from $b=7$\,fm Au+Au collisions at 
four fixed values of $p_T$ (in GeV/$c$), as indicated.
}
\label{F5} 
\end{figure}
%

\begin{figure}
\begin{center}
\includegraphics[height=0.7\linewidth,angle=-90,clip=]{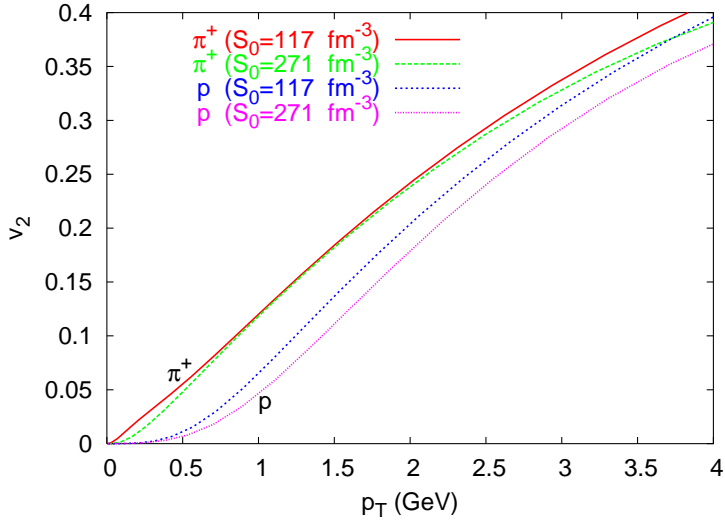}
\end{center}
\caption{(Color online) Differential elliptic flow $v_2(p_T)$ 
for thermal pions and protons from $b=7$\,fm Au+Au collisions at RHIC 
($s_0=117$\,fm$^{-3}$) and LHC ($s_0=271$\,fm$^{-3}$) energies,
plotted together for comparison.}
\label{F6} 
\end{figure}

This is explored more quantitatively in the lower panel of Fig.~\ref{F5}
which shows the differential elliptic flow of directly emitted pions at 
fixed $p_T$ as a function of $\sqrt{s}$ (using the fit in Fig.~\ref{F1}
to translate $s_0$ into $\sqrt{s}$), for four different $p_T$ values.
The three curves corresponding to $p_T=0.1,\ 0.5$ and 1\,GeV/$c$ all show
this non-monotonic behavior of a $v_2(p_T)$ that first rises and then 
decreases with collision energy. The increase at low $\sqrt{s}$ values 
is easy to understand: at low collision energies, the fireball decouples 
before the elliptic flow can saturate; by increasing the collision energy,
the fireball lifetime is increased, allowing the elliptic flow to grow and
reach values closer to its asymptotic saturation value. Between SPS and 
RHIC energies, the fireball lifetime is sufficiently long for the elliptic 
flow to more or less saturate \cite{Kolb:1999it}. The explanation for 
the decrease of $v_2(p_T)$ at higher energies is more subtle: As the 
collision energy increases further, the magnitude of the {\em radial} 
flow continues to grow, too, leading to ever flatter $p_T$-spectra. 
Flatter $p_T$-spectra show less azimuthal variation and thus exhibit 
smaller $v_2$ coefficients \cite{Huovinen:2001cy}. This flattening effect 
of radial flow on the $p_T$-spectra is particularly pronounced at low 
$p_T$ (see Fig.~\ref{F2}). This may explain why the non-monotonic beam 
energy dependence of $v_2(p_T)$ is only seen at low $p_T$ up to about 
1\,GeV/$c$ and seems to disappear above $p_T=2$\,GeV/$c$ (dashed line 
in the lower panel of Fig.~\ref{F5}). We expect the non-monotonicity 
of $v_2(p_T)$ to persist to larger $p_T$
values for protons whose low-$p_T$ spectra are more strongly affected 
by radial flow than those of pions (see Fig.~\ref{F2}). Figure~\ref{F6}
supports this expectation by showing that for protons (in contrast to 
pions) the differential elliptic flow at the LHC is {\em smaller} than at 
RHIC over the entire $p_T$-range shown in that Figure.

%
%

It is important to note, however, that, while the differential elliptic 
flow at fixed  $p_T$ {\em decreases} between RHIC and LHC energies, the 
total momentum anisotropy, reflected by the $p_T$-integrated elliptic
flow coefficient $v_2$, {\em increases} from RHIC to LHC. Radial flow 
flattens the LHC spectra dramatically, putting a larger weight on the 
larger $v_{2}$ values at higher $p_{T}$, and as a result the 
{\em $p_{T}$-integrated} elliptic flow is larger at the LHC, even though 
at any fixed $p_T$ value below 1 GeV/$c$ it is smaller than at RHIC. This 
is shown in the upper panel of Fig.~\ref{F5} for both directly emitted pions 
(dashed line) and all pions including resonance decay contributions 
(solid line). Again, one sees first a rise of $v_2$ at low $\sqrt{s}$ (AGS
energies), followed by a decrease between SPS and RHIC and another rise 
between RHIC and LHC \cite{Kolb:1999it}. The non-monotonicity of the
{\em $p_T$-integrated} $v_2$ can be related to the quark-hadron phase
transition and the corresponding softening of the EOS in the transition
region \cite{Kolb:1999it}. In the experimental data this phase transition
signature is, unfortunately, washed out by strong viscous effects in the
late hadronic stage of the fireball expansion which become increasingly 
important at low collision energies \cite{Teaney:2001av,Hirano:2005xf,%
Heinz:2004ar}(where the fireball spends most of its time in the hadronic 
phase). As a result, the experimentally measured integrated elliptic flow 
$v_2$ rises monotonically with $\sqrt{s}$, approaching the ideal fluid
limit only at or above RHIC energies \cite{Hirano:2007xd}.

The upper panel of Fig.~\ref{F5} shows that at all collision energies the 
directly thermally emitted pions show more $p_{T}$-integrated elliptic 
flow than all pions together. From this we can deduce that pions emitted 
from resonance decays have a lower momentum anisotropy. This is at least 
partially understood by the fact that decay pions typically have smaller 
transverse momenta than their parent resonances \cite{Concepts}, and that 
the parent resonances, being heavier, contribute less elliptic 
flow at low $p_T$ than pions (see Fig.~\ref{F6}) \cite{Hirano:2000eu}.

Comparison of the curves shown in the upper panel of Fig.~\ref{F5}
with those in the right panel of Fig.~2 in Ref.~\cite{Hirano:2007xd} 
shows that in a hybrid model approach (which includes viscous effects in 
the hadronic phase) the $p_T$-integrated elliptic flow increases by 
about 25\% between RHIC and LHC whereas in the ideal fluid approach (which
neglects hadronic viscosity) it increases by less than 10\%. The
largest contribution to the expected increase of $v_2$ from RHIC to 
LHC is therefore due to the {\em disappearance} of late hadronic viscous 
effects between RHIC and LHC because the fireball spends a smaller fraction
of it time in the hadronic phase and the elliptic flow saturates before
the QGP converts to hadrons. 

\section{Conclusions}
\label{sec5}

We have shown that, while the radial flow and the $p_T$-integrated 
elliptic flow {\em increase} from RHIC to LHC, the differential 
elliptic flow at fixed $p_T<1.5$\,GeV/$c$ {\em decreases} in the same 
collision energy range. This decrease of $v_2(p_T)$ is driven by a 
(relative) depletion of low $p_T$ hadrons by radial flow which pushes 
the hadrons to larger transverse momenta. The observed ``saturation'' 
of $v_2(p_T)$ seen by PHENIX \cite{Adler:2004cj} between $\sqrt{s}=63$ 
and 200 GeV likely signals the onset of this kinematic effect and should 
thus not be interpreted without further scrutiny as a signature for the 
quark-hadron phase transition.  

When experimental data will become available for very high energy 
collisions at the LHC, the ideal fluid dynamical calculations presented
here can serve as a benchmark for comparison with experiment, providing 
insights into the validity of ideal hydrodynamics at high energy densities 
and permitting quantification of deviations from ideal fluid behaviour. 
This should facilitate the discovery of possible novel effects, beyond 
those expected from collective dynamics, at LHC energies.

{\bf Acknowledgements:}
We thank Rupa Chatterjee, Evan Frodermann, Richard Furnstahl and Huichao 
Song for helpful comments. The work of G.K. was supported by the National 
Science Foundation under grant PHY-0354916, that of U.H. by the U.S. 
Department of Energy under contract DE-FG02-01ER41190.

%

\end{document}